\documentstyle[prc,aps,epsfig]{revtex}

\def\r{\mbox{\boldmath $r$}}
\def\R{\mbox{\boldmath $R$}}
\def\p{\mbox{\boldmath $p$}}

\def\q{\mbox{\boldmath $q$}}

\def\J{\mbox{\boldmath $J$}}

\def\ss{\mbox{\boldmath $\sigma$}}

\begin{document}
\draft
\title{Short-range and tensor correlations in the $^{16}$O(e,e$'$pn) reaction}
\author{C.~Giusti$^1$, H.~M\"uther$^2$, F.~D.~Pacati$^1$ and M.~Stauf$^2$}
\address{$^1$Dipartimento di Fisica Nucleare e Teorica dell'Universit\`a, 
Pavia\\
and Istituto Nazionale di Fisica Nucleare, Sezione di Pavia, Italy\\
$^2$Institut f\"ur Theoretische Physik, Universit\"at T\"ubingen,\\
Auf der Morgenstelle 14, D-72076 T\"ubingen, Germany}
\date{\today}
\maketitle

\begin{abstract}%
The cross sections for electron induced two-nucleon knockout reactions are 
evaluated for the example of the $^{16}$O(e,e$'$pn)$^{14}$N reaction leading 
to discrete states in the residual nucleus $^{14}$N. These calculations account
for the effects of nucleon-nucleon correlations and include the contributions of
two-body meson exchange currents as the pion seagull, pion in flight and the
isobar current contribution. The effects of short-range as well as tensor
correlations are calculated within the framework of the coupled cluster method
employing the Argonne V14 potential as a model for a realistic nucleon-nucleon
interaction. The relative importance of correlation effects as compared to the
contribution of the meson exchange currents depends on the final state of the
residual nucleus. The cross section leading to specific states, like e.g. the
ground state of $^{14}$N, is rather sensitive to the details of the correlated
wave function. 
\end{abstract}

\section{Introduction}

The nuclear shell-model, describing a nucleus as a system of nucleons moving
in a mean field, describes many basic features of nuclear structure.
Nevertheless, it is well known, that the strong components of a realistic model
for the nucleon-nucleon (NN) interaction induce correlations into the nuclear
wave function, which are beyond this mean field description. In particular one
has to consider the two-nucleon short-range correlations, induced by the 
repulsive components of the NN interaction, and the tensor correlations which
are mainly due to the strong tensor components of the pion exchange contribution
to the NN interaction. A detailed investigation of these correlations should
provide insight into the structure of the interaction of two nucleons in a
nuclear medium. Therefore it has always been a great challenge of nuclear
physics to develop experiments and theoretical models, which explore these
correlations.

Photoinduced two-nucleon
knockout experiments like ($\gamma$,NN) or (e,e$'$NN) reactions seem to be
a powerful tool for such investigations since the probability that a real or
virtual photon is absorbed by a pair of nucleons should be a direct measure for
the correlations between these nucleons. Such triple coincidence experiments
have been made possible by the progress in accelerator and detector
technology and first measurements of the exclusive $^{16}$O(e,e$'$pp)$^{14}$C 
reaction have been performed at NIKHEF in Amsterdam~\cite{Gerco,NIKHEF} and MAMI
in Mainz~\cite{Rosner}. First investigations on these
data~\cite{Gerco,NIKHEF,Rosner,klas1} indicate that resolution of discrete 
final states provides an interesting tool to disentangle and thus separately 
investigate contributions of one-body currents, due to short-range 
correlations, and two-body isobar currents. In particular, clear signatures for
short-range correlations have been obtained for the transition to the ground
state of $^{14}$C. This result gives rise to the hope that a similar separation
between two-body currents and correlations is possible also in the (e,e$'$pn) 
reaction. This would be of particular interest as tensor correlations will 
predominantly be present in the wave function of a $pn$ pair. 
Thus (e,e$'$pn) experiments would complete the information on NN correlations.
It is obvious that such experiments are more difficult as they require a triple
coincidence measurement with a detection of a neutron. A proposal for the
experimental study of the exclusive $^{16}$O(e,e$'$pn)$^{14}$N reaction has
been recently approved in Mainz~\cite{MAMI}. This first experiment will 
demonstrate the feasibility of triple coincidence experiments including
neutrons with high resolution and will serve as exploratory study for the
investigations of $np$ correlations.

Also the theoretical analysis of the $pn$ data is a bit more involved than the
analysis of the $pp$ data since the contribution of the correlations to the
$pn$ knockout has to compete with a significant contribution from pion exchange
currents~\cite{Oxford,mec1,mec2,Peccei,gnn,vander,mache,ryck,ic}. Furthermore, 
one has to treat the tensor correlations appropriately. It is the aim of this
investigation to employ correlated many-body wave functions, which are derived
from a realistic NN interaction, in a calculation of (e,e$'$pn) cross sections 
which accounts for the contribution of one- and two-body parts in the nuclear
current operator.

The correlated wave functions are determined in the framework of the coupled
cluster method, which is also often referred to as the $\exp(S)$ 
method~\cite{kuem,zab1,bish,bogd,stauf}, using the so-called $S_2$ 
approximation. 
This approach is similar to the evaluation of the two-body spectral function in
terms of the Brueckner $G$-matrix as it has been used e.g.~in the analysis of
the $^{16}$O(e,e$'$pp)$^{14}$C reaction~\cite{NIKHEF,klas1,klas2}. The Argonne 
V14 potential~\cite{v14} has been used for the NN interaction.

After this introduction we will shortly review the coupled cluster method in
section 2 and discuss some features of the correlated wave functions derived 
in this scheme. The calculation of the matrix elements for the one- and two-body
current contributions and the resulting cross sections are described in section
3. After the discussion of the numerical results for the 
$^{16}$O(e,e$'$pn)$^{14}$N reaction in section 4, we will summarize our main
results and conclusions.


\section{Correlated NN wave functions in the coupled cluster method}

The basic features of the coupled cluster method have been described already in
the review article by K\"ummel et al.~\cite{kuem}. More recent developments and
applications can be found in~\cite{bish}. Here we will only present some basic
equations, which are necessary to describe the approach we have used to
determine the correlated nuclear wave functions to be used in the calculation of
cross sections for the exclusive $^{16}$O(e,e$'$pn)$^{14}$N reaction. The
many-body wave function of the coupled cluster or $\exp(S)$ method can be 
written
\begin{equation}
\vert \Psi > =  \exp \left(\sum_{n=1}^A \hat S_n\right) \vert \Phi >\, .
\label{eq:exps}
\end{equation}
The state $\vert \Phi >$ refers to the uncorrelated model state, which we have
chosen to be a Slater determinant of harmonic oscillator functions with an
oscillator length $b$=1.72 fm, which is appropriate for the description of our
target nucleus $^{16}$O. This choice leads to small amplitudes for the
operator $S_1$. The linked $n$-particle $n$-hole excitation operators
can be written
\begin{equation}
\hat S_n = \frac{1}{n!^2} \sum_{\nu_i\rho_i} <\rho_1\dots\rho_n\vert S_n \vert
\nu_1\dots \nu_n> a_{\rho_1}^\dagger \dots  a_{\rho_n}^\dagger a_{\nu_n} \dots
a_{\nu_1}\,.\nonumber
\end{equation}
Here and in the following the sum is restricted to oscillator states $\rho_i$ 
which are unoccupied in
the model state $\vert \Phi >$, while states $\nu_i$ refer to states which are
occupied in $\vert \Phi >$. For the application discussed here we assume the
so-called $S_2$ approximation, i.e.~we restrict the correlation operator in
(\ref{eq:exps}) to the terms with $\hat S_1$ and $\hat S_2$. One may introduce
one- and two-body wave functions
\begin{eqnarray}
\psi_1 \vert \nu_1> & = & \vert \nu_1 > + \hat S_1 \vert \nu_1 > \nonumber \\
\psi_2 \vert \nu_1 \nu_2 > & = & {\cal A} \, \psi_1 \vert \nu_1> \psi_1 \vert
\nu_2> + \hat S_2 \vert \nu_1 \nu_2 > \label{eq:psin}
\end{eqnarray}
with ${\cal A}$ denoting the operator antisymmetrizing the product of one-body
wave functions.
Using these definitions one can write the coupled equations for the evaluation
of the correlation operators $\hat S_1$ and $\hat S_2$ in the form
\begin{equation}
<\alpha \vert \hat T_1 \psi_1 \vert \nu > + \sum_{\nu_1} <\alpha \nu_1 \vert
\hat T_2 \hat S_2 + \hat V_{12} \vert \nu \nu_1 > = \sum_{\nu_1} 
\epsilon_{\nu_1\nu} <\alpha \vert \psi_1 \vert \nu_1 >\,, \label{eq:hf}
\end{equation} 
where $\hat T_i$ stands for the operator of the kinetic energy of particle $i$ 
and $\hat V_{12}$ is the two-body potential. Furthermore we introduce the
single-particle energy matrix defined by
\begin{equation}
\epsilon_{\nu_1\nu} = <\nu_1 \vert \hat T_1 \vert \nu > + \sum_{\nu'} < \nu_1
\nu' \vert \hat V_{12} \psi_2 \vert \nu \nu'> \nonumber
\end{equation}
The Hartree-Fock type equation (\ref{eq:hf}) is coupled to a two-particle
equation of the form
\begin{eqnarray}
0 & = & <\alpha \beta \vert \hat Q \left[ (\hat T_1 + \hat T_2)\hat S_2 + 
\hat
V_{12}\psi_2 + \hat S_2 \hat P \hat V_{12} \psi_2 \right] \vert \nu_1 \nu_2 >
\nonumber \\
&& \qquad  
- \sum_{\nu} \left( <\alpha \beta \vert \hat S_2 \vert \nu \nu_2 >
\epsilon_{\nu\nu_1} + <\alpha \beta \vert \hat S_2 \vert \nu_1 \nu >
\epsilon_{\nu\nu_2} \right)  \label{eq:s2}
\end{eqnarray}
In this equation we have introduced the Pauli operator $\hat Q$ projecting on
two-particle states, which are not occupied in the uncorrelated model state
$\vert \Phi >$ and the projection operator $\hat P$, which projects on
two-particle states, which are occupied. If for a moment we ignore the term in
(\ref{eq:hf}) which is represented by the operators $\hat T_2 \hat S_2$ and
also the term in (\ref{eq:s2}) characterized by the operator  $\hat S_2 \hat P 
\hat V_{12}$ the solution of these coupled equations corresponds to the
Brueckner-Hartree-Fock approximation and we can identify the matrix elements of 
$\hat V_{12} \psi_2$ with the Brueckner $G$-matrix. Indeed the effects of these
two terms are rather small and we have chosen the coupled cluster approach
mainly because it provides directly correlated two-body wave functions.

The coupled equations (\ref{eq:hf}) and (\ref{eq:s2})  have been solved by
expanding the amplitudes defining $S_2$ and the wave functions $\psi_2$ in terms
of product wave functions for the relative and center of mass coordinates  
\begin{equation}
\r_{12}= \r_{1}- \r_{2}, \, \, \, \, \,\R= \frac{\r_{1} +\r_{2}}{2}
\end{equation}
The harmonic oscillator wave functions corresponding to the model state $\vert
\Phi >$  turn out to provide a useful basis for the center of mass wave
functions, while the relative wave functions are represented more efficiently in
a complete basis of eigenstates for a spherical box of radius $r_{\mbox{max}}$. 
It turns out that the results are independent on the choice of $r_{\mbox{max}}$
if this value is chosen large enough as compared to the relative distances
between two nucleons in $^{16}$O ($r_{\mbox{max}}\approx$ 10 fm or
larger)~\cite{stauf}. 

This means that we can write the correlated two-body wave functions with
${\ss}_i$ denoting the spin and $J$  referring to the total angular momentum of
the pair
\begin{eqnarray}
<\r_{12},\R ,{\ss}_1,{\ss}_2\vert \psi_2 \vert \nu_1 \nu_2 >_J & = & 
\sum_{lSjNL} \,c_{lSjNL}^{\nu_1 \nu_2J} \, \phi_{lSjNL}^{\nu_1 \nu_2J}(r_{12}) 
R_{NL}(R)  \nonumber \\ & &\qquad \times  \left[\Im ^{j}_{lS}
(\Omega_r,{\ss}_1,{\ss}_2) \,
Y_{L}(\Omega_R)\right]^{J},
\label{eq:ppover1}
\end{eqnarray}
If we would ignore all correlation effects, this equation corresponds to the
Talmi-Moshinsky transformation of harmonic oscillator states from the laboratory
frame (oscillator states $\nu_1$ and $\nu_2$ ) to relative and center of mass
coordinates. In this case the eq.~(\ref{eq:ppover1}) is simplified. The relative
wave functions $\phi_{lSjNL}^{\nu_1 \nu_2J}$ do not depend on the quantum 
numbers $S,j,N,L,\nu_1$, $\nu_2$ and $J$ and are simply given in terms of 
harmonic oscillator functions. The coefficients $c_{lSjNL}^{\nu_1 \nu_2J}$ are 
the well known coefficients of the Talmi-Moshinsky transformation and are 
different from zero only if the oscillator relation
$$
2n + l\, +\, 2N+L \, = \, 2n_1+l_1\,+\,2n_2+l_2
$$
holds, with $n_i,l_i$ referring to oscillator quantum numbers of state $\nu_i$.

\begin{figure}
\epsfysize=12.0cm
\begin{center}
\makebox[16.4cm][c]{\epsfbox{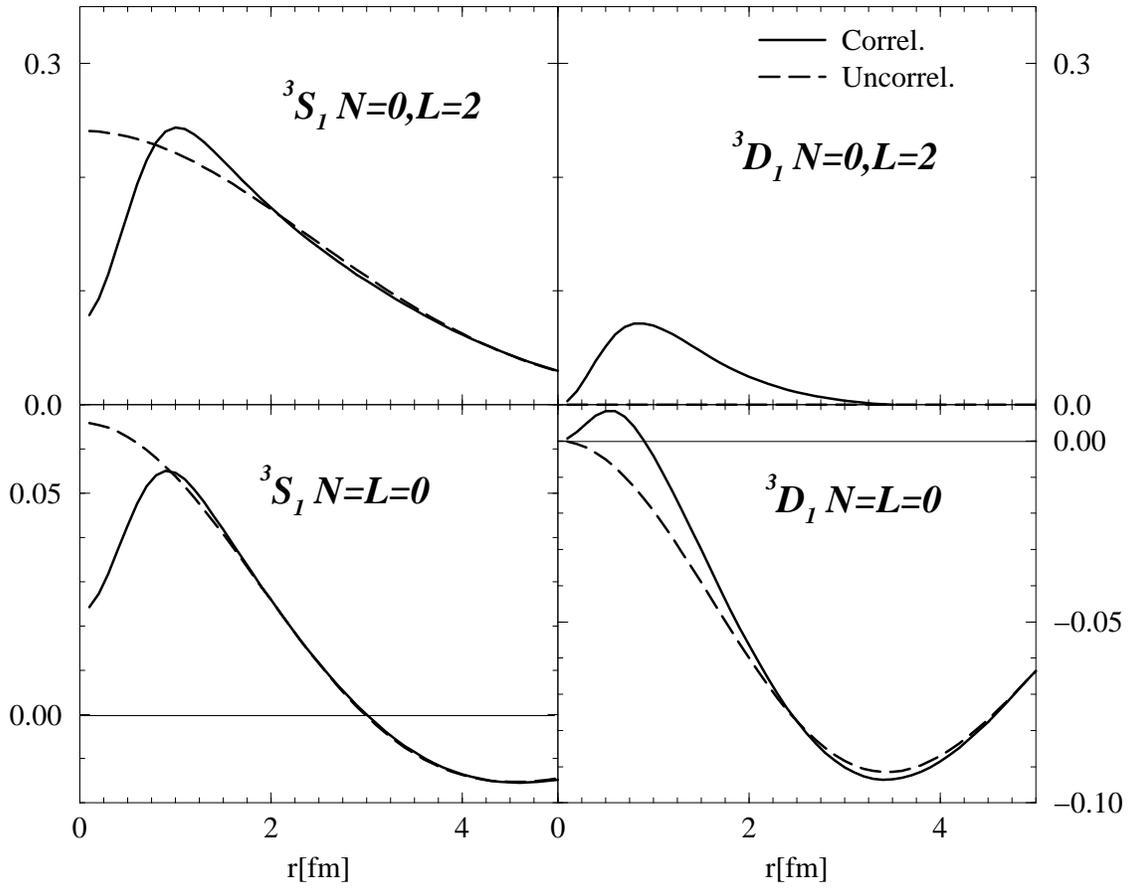}}
\end{center}
\caption[]{Relative wave functions for correlated and uncorrelated two-body wave
functions in the case of a proton and a neutron in the $p_{1/2}$ shell coupled
to $J=1$. Further discussion in the text.\label{fig:mosh}}
\end{figure}

In order to visualize the effects of NN correlations we would like to discuss a
specific example, a proton and a neutron in $p_{1/2}$ shell coupled to $J=1$.
The non-vanishing Talmi-Moshinsky transformation coefficients are given by
\begin{eqnarray}
\sqrt{\frac{1}{54}} & , & \mbox{for }^3S_1,\,N=L=0\nonumber\\
-\sqrt{\frac{10}{27}} & , & \mbox{for }^3D_1,\,N=L=0\nonumber\\
\sqrt{\frac{10}{27}} & , & \mbox{for }^3S_1,\,N=0,L=2\nonumber\\
-\sqrt{\frac{1}{54}} & , & \mbox{for }^3S_1,\,N=1,L=0\nonumber\\
\sqrt{\frac{6}{27}} & , & \mbox{for }^1P_1,\,N=0,L=1 . \label{eq:mosh}
\end{eqnarray}
The relative wave functions, calculated with the Argonne V14 potential, for 
some of these channels are displayed in
Fig.~\ref{fig:mosh}. In the case of the uncorrelated approach (dashed lines)
these wave functions are simply given by the corresponding harmonic oscillator
functions multiplied with one of the Talmi-Moshinsky transformation coefficients
listed in (\ref{eq:mosh}). The correlated wave functions, however, show
different shapes. At small relative distances $r$ the amplitudes for the $^3S_1$
partial waves are typically 
smaller than for the corresponding uncorrelated wave, reflecting the repulsive
interaction at such small distances. For larger distances the correlated wave
function approach the uncorrelated ones (healing property). Note, that
correlated wave functions are also obtained for channels like the
$^3D_1,\,N=0,L=2$, for which the uncorrelated wave function vanishes. These
are mainly due to the tensor correlations. It is also worth mentioning that our
approach yields state-dependent correlation functions in the sense that the
correlated wave functions $\phi_{lSjNL}^{\nu_1 \nu_2J}$ depend on all quantum
numbers which are attached as sub- or superscript. This state dependence can be
seen in the examples displayed in Fig.~\ref{fig:mosh}.

\section{Cross section and reaction mechanism}

The coincidence cross section for the reaction induced by an electron 
with momentum $\p_{0}$ and energy $E_{0}$, with $E_{0}=|\p_{0}|=p_{0}$, where 
two nucleons, with momenta $\p'_{1}$, and $\p'_{2}$ 
and energies $E'_{1}$ and $E'_{2}$, are ejected from a nucleus is given, in the
one-photon exchange approximation and after 
integrating over $E'_{2}$, by~\cite{Oxford,GP} 
\begin{equation}
\frac{{\mathrm d}^{8}\sigma}{{\mathrm d}E'_{0}{\mathrm d}\Omega
{\mathrm d}E'_{1}{\mathrm d}\Omega'_{1} 
{\mathrm d}\Omega'_{2}} = K \Omega_{\mathrm f} f_{\mathrm{rec}} 
|j_\mu J^\mu|^2 .
\label{eq:cs}
\end{equation}
In Eq.~(\ref{eq:cs}) $E'_{0}$ is the energy of the scattered electron with
momentum $\p'_{0}$, $K = e^4{p'_{0}}^2/4\pi^2 Q\,^4$ where 
$Q^2 = \q\,^2 - \omega^2$, with $\omega = E_{0} - E'_{0}$ and 
$\q = \p_0 - \p'_0$, is the four-momentum transfer. The quantity
$\Omega_{\mathrm f} = p'_{1} E'_{1} p'_{2} E'_{2}$ is the phase-space 
factor and integration over $E'_{2}$ produces the recoil factor 
\begin{equation}
f_{\mathrm{rec}}^{-1} = 1 - \frac{E'_{2}}{E_{\mathrm B}} \, \frac{\p'_{2}\cdot 
\p_{\mathrm B}}{|\p'_{2}|^2},
\end{equation}
where $E_{\mathrm B}$ and $\p_{\mathrm B}$ are the energy and momentum of the 
residual nucleus. The cross section is given by the square of the scalar
product of the relativistic electron current $j^\mu$ and of the nuclear 
current $J^\mu$, which is given by the Fourier transform of the transition 
matrix elements  of the charge-current density operator between initial and 
final nuclear states
\begin{equation}
J^\mu (\q) = \int < \Psi_{\mathrm{f}} | \hat{J}^\mu(\r) |\Psi_{\mathrm{i}} >
{\mathrm{e}}^{\,{\mathrm{i}}{\footnotesize \q} \cdot
{\footnotesize \r}} {\mathrm d}\r.
\label{eq:jm}
\end{equation} 

If the residual nucleus is left in a discrete eigenstate of its Hamiltonian,
i.e. for an exclusive process, and under the assumption of a direct knockout
mechanism, Eq.~(\ref{eq:jm}) can be written as~\cite{GP,eepp} 
\begin{eqnarray}
J^\mu(\q) & = &  \int 
\psi_{\mathrm{f}}^{*}(\r_{1}\ss_{1},\r_{2}\ss_{2})
J^\mu(\r,\r_{1}\ss_{1},\r_{2}\ss_{2})\psi_{\mathrm{i}}
(\r_{1}\ss_{1},\r_{2}\ss_{2}) \nonumber \\
& & \times \,{\mathrm{e}}^{\,{\mathrm{i}}{\footnotesize \q} \cdot
{\footnotesize \r}} {\mathrm d}\r{\mathrm d}\r_{1} {\mathrm d}\r_{2}
{\mathrm d}\ss_{1} {\mathrm d}\ss_{2} . \label{eq:jq}
\end{eqnarray}

Eq.~(\ref{eq:jq}) contains three main ingredients: the two-nucleon overlap
integral $\psi_{\mathrm{i}}$, the nuclear current $J^\mu$ and the final-state
wave function $\psi_{\mathrm{f}}$. 

In the model calculations the final-state wave function $\psi_{\mathrm {f}}$ 
includes the interaction of each one of the two outgoing nucleons with the 
residual nucleus while their mutual interaction is neglected. Therefore, the 
scattering state is written as the product of two uncoupled single-particle 
distorted wave functions, eigenfunctions of a complex phenomenological optical 
potential which contains a central, a Coulomb and a spin-orbit term. 

The nuclear current operator in Eq.~(\ref{eq:jq}) is the sum of a one-body 
and a two-body part. In the one-body part convective and spin currents are 
included. The two-body current is derived from the effective Lagrangian of 
ref.~\cite{Peccei}, performing a non relativistic reduction of the 
lowest-order Feynman diagrams with one-pion exchange. We have thus currents 
corresponding to the seagull and pion-in-flight diagrams and to the diagrams 
with intermediate isobar configurations~\cite{gnn}, i.e.

\begin{eqnarray}
\J^{(2)}(\r,\r_{1}\ss_{1},\r_{2}\ss_{2}) & = &
\J^{\mathrm{sea}}(\r,\r_{1}\ss_{1},\r_{2}\ss_{2}) +
\J^{\pi}(\r,\r_{1}\ss_{1},\r_{2}\ss_{2}) \nonumber \\
& + & \J^{\Delta}(\r,\r_{1}\ss_{1},\r_{2}\ss_{2}) . \label{eq:nc}
\end{eqnarray}

Details of the nuclear current components and the values of the parameters 
used in the calculations are given in Ref.~\cite{gnn}.

The two-nucleon overlap integral $\psi_{\mathrm{i}}$ contains the information 
on nuclear structure and allows one to write the cross section in terms of the 
two-hole spectral function~\cite{Oxford}. 
For a discrete final state of the $^{14}$N nucleus, with angular momentum
quantum number $J$, the state $\psi_{\mathrm{i}}$ is expanded in terms of the
two-hole wave functions defined in (\ref{eq:ppover1}) as
\begin{eqnarray}
\psi_{\mathrm{i}}^{J}(\r_{1}{\ss}_{1},
{\r}_{2}{\ss}_{2}) & = & \sum_{\nu_1 \nu_2}
 a^{J}_{\nu_1\nu_2} <\r_{12},\R ,{\ss}_1,{\ss}_2\vert \psi_2 \vert \nu_1 \nu_2 >
\label{eq:ppover}
\end{eqnarray}
The expansion coefficients $a^{J}_{\nu_1\nu_2}$ are determined from a
configuration mixing calculation of the two-hole states in $^{16}$O, which can
be coupled to the angular momentum and parity of the requested state. The
residual interaction for this shell-model calculation is also derived from the
Argonne V14 potential and corresponds to the Brueckner G-matrix. The expansion
coefficients $a^{J}_{\nu_1\nu_2}$, given in Table I for the transitions to the
low-lying states of $^{14}$N considered in this paper, are renormalized to 
account for the spectroscopic factors of the single-particle states as it is 
described in \cite{klas2}.


\section{The $^{16}$O(\lowercase{e,e$'$pn})$^{14}$N knockout
reaction to discrete final states}

In this section we present numerical results of the cross section of the 
$^{16}$O(e,e$'$pn)$^{14}$N reaction for transitions to the lowest-lying
discrete states in the residual nucleus that are expected to be strongly
populated by direct knockout. We consider three states, all of them with 
positive parity and $T=0$: the 1$^+_1$ ground state of $^{14}$N, the 1$^+_2$ 
state at 3.95 MeV and the 2$^+$ state at 7.03 MeV. These states are of
particular interest for our investigation since they can be separated in 
high-resolution experiments such as that recently proposed at MAMI~\cite{MAMI}.
The configuration mixing calculations described at the end of section 3
predict excitation energies of 4.54 MeV for the 1$^+_2$ and 7.17 MeV for the
2$^+$ state.

In the calculations each state is characterized by a particular value of the
missing energy, given by
\begin{equation}
E_{2\mathrm{m}} = \omega - T'_{1} - T'_{2} -T_{\mathrm{B}} =
E_{\mathrm{s}} + E_{\mathrm{x}} , \label{eq:em}
\end{equation}
where $T'_{1}$, $T'_{2}$ and $T_{\mathrm{B}}$ are the kinetic energies of the
two outgoing nucleons and of the residual nucleus, respectively,
$E_{\mathrm{s}}$ is the separation energy at threshold for two-nucleon
emission and $E_{\mathrm{x}}$ is the excitation energy of the residual nucleus.
 
As an example, we have performed calculations in the so-called super-parallel 
kinematics~\cite{GP}, where the knocked-out nucleons are detected parallel and 
anti-parallel to the transferred momentum $\q$. In this kinematics, for a fixed
value of the energy and momentum transfer and for a particular final state, it
is possible to explore, for different values of the kinetic energies of the 
outgoing nucleons, all possible values of the recoil ($p_{\mathrm{B}}$) or 
missing momentum ($p_{2{\mathrm{m}}}$) distribution, where 
\begin{equation}
\p_{2\mathrm{m}} = \p_{\mathrm{B}} =  \q -\p'_{1} - \p'_{2}.  \label{eq:pm}
\end{equation}

The super-parallel kinematics is favored by the fact that only two structure
functions, the longitudinal and transverse structure functions, contribute to 
the cross section, as in the inclusive electron scattering, and, as in that 
case, can in principle be separated by a Rosenbluth plot~\cite{GP}. This 
kinematical setting is also favorable from the experimental point of view. It 
has been realized in a recent $^{16}$O(e,e$'$pp)$^{14}$C experiment at 
MAMI~\cite{Rosner} and has been proposed for the first experimental study of 
the $^{16}$O(e,e$'$pn)$^{14}$N reaction~\cite{MAMI}. The choice of the same 
kinematics is of particular interest for the comparison of cross sections and 
reaction mechanisms for $pp$ and $pn$ emission, for the investigation of the 
relative strength of $pp$ and $pn$ correlations and of their contributions in 
the two processes and for the determination of both types of correlations.  

The same kinematical parameters of the MAMI experiment have been adopted in the
calculations, with $E_0 =855$ MeV, electron scattering angle $\theta= 
18^{\mathrm{o}}$, $\omega = 215$ MeV and $q= 316$ MeV/$c$. The proton is 
emitted parallel and the neutron antiparallel to the momentum transfer $\q$. 
This choice appears well suited to reduce the contribution of meson-exchange
currents and emphasize effects due to correlations~\cite{KM}.

\begin{figure}
\epsfysize=9.0cm
\begin{center}
\makebox[16.4cm][c]{\epsfbox{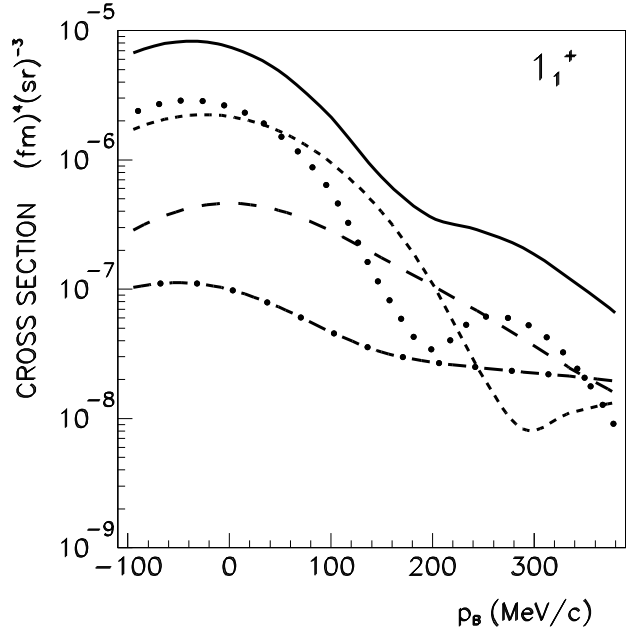}}
\end{center}
\caption[]{The differential cross section  of the $^{16}$O(e,e$'$pn)
reaction as a function of the recoil momentum $p_{\mathrm{B}}$ for the 
transition to the $1^+_1$ ground state of $^{14}$N ($E_{2\mathrm{m}} = 22.96$ 
MeV), in the super-parallel kinematics with $E_{0} = 855$ MeV, and 
$\omega = 215$ MeV $q = 316$ MeV/$c$. The recoil-momentum distribution is 
obtained changing the kinetic energies of the outgoing nucleons. Separate 
contributions of the one-body, seagull, pion-in-flight and $\Delta$-current are
shown by the dotted, short-dashed, dot-dashed and long-dashed lines, 
respectively. The solid line gives the total cross section. The optical 
potential is taken from Ref.~\cite{Nad}. Positive (negative) values of 
$p_{\mathrm{B}}$ refer to situations where 
${\mbox{\boldmath $p$}}_{\mathrm{B}}$ is parallel (antiparallel) to 
${\mbox{\boldmath $q$}}$.
\label{fig:fig1}
}
\end{figure}

The cross section for the transition to the 1$^+_1$ ground state of $^{14}$N is
shown in Fig.~\ref{fig:fig1}. Separate contributions of the different terms of
the nuclear current are shown in the figure and compared with the total cross
section. The contribution of the one-body current, entirely due to 
correlations, is large. It is of the same size as that of the pion seagull 
current. The contribution of the $\Delta$-current is much smaller at lower 
values of $p_{\mathrm{B}}$, whereas for values of $p_{\mathrm{B}}$ larger than 
100 MeV/$c$ it becomes comparable with that of the other components. The 
contribution of the pion-in-flight current is generally much smaller, but at 
higher values of the recoil momentum. The final result indicates that all the 
terms of the nuclear current give a significant contribution to the calculated 
cross section and that interference effects are not too large. The major role 
is played in the considered situation by the 1-body and seagull terms. In 
particular, the one-body current and thus correlations yield a contribution 
that is important to determine the size and the shape of the cross section.

The shape of the recoil-momentum distribution is driven by the c.m. orbital
angular momentum $L$ of the knocked out pair. This feature, that is fulfilled 
in a factorized approach~\cite{Got}, where final-state interaction is neglected
and $\p_{\mathrm{B}}$ is opposite to the total momentum of the initial nucleon
pair, is not spoiled in an unfactorized approach by final-state
interaction~\cite{klas1,gnn,eepp}. 

\begin{figure}
\epsfysize=9.0cm
\begin{center}
\makebox[16.4cm][c]{\epsfbox{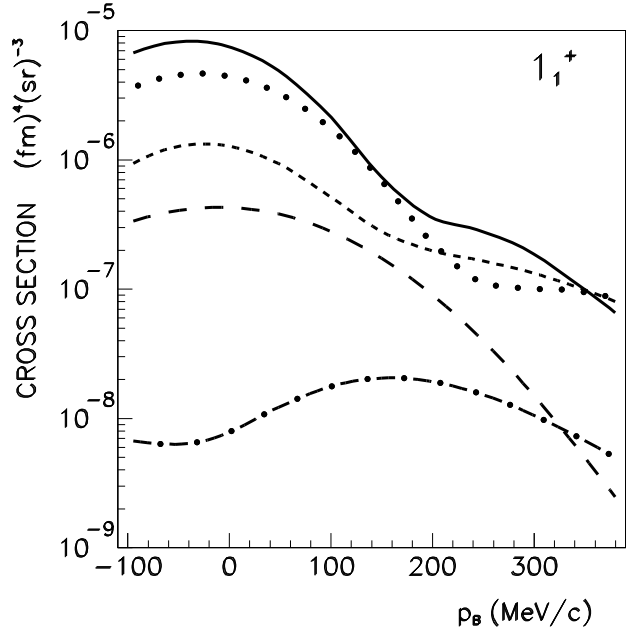}}
\end{center}
\caption[]{The differential cross section  of the $^{16}$O(e,e$'$pn)
reaction as a function of the recoil momentum $p_{\mathrm{B}}$ for the same 
transition and in the same kinematics as in Fig.~\ref{fig:fig1}. Optical
potential as in Fig.~\ref{fig:fig1}. Separate contributions of different
partial waves of relative motions are drawn: dotted line for $^3S_1$, 
dot-dashed line for $^1P_1$, short-dashed line gives the $^3D_1$ component
already present in the uncorrelated wave function and long-dashed line the 
$^3D_1$ component due to tensor correlations. The 
solid line is the same as in Fig.~\ref{fig:fig1}. 
\label{fig:fig2}
}
\end{figure}

Different partial waves of relative and c.m. motion contribute to the
two-nucleon overlap function in Eqs.~(\ref{eq:ppover}) and (\ref{eq:ppover1}). 
Each transition is thus
characterized by different components, with specific values of $L$ (see Section
2), whose relative weights will determine the shape of the recoil momentum 
distribution. This can be seen in Fig.~\ref{fig:fig2}, where the separate 
contributions of the different partial waves of relative motion are displayed. 
For the transition to the $1^+_1$ ground state there are the following relative
states: $^3S_1$, combined with a c.m. $L= 0$ and $L= 2$, $^1P_1$, combined with
$L= 1$, and $^3D_1$. For the $^3D_1$ relative wave function we 
have separated in the figure the component already present in the uncorrelated 
wave function, which is combined with $L= 0$, and the component produced by 
tensor correlations and not present in the uncorrelated wave function, which is 
combined with $L= 0$ and $L= 2$ (see also Fig.~\ref{fig:mosh} and discussion
above). In the following we will call these two terms 
 $^3D_1$ and $^3D_1^{\mathrm T}$, respectively. The contribution of the  
$^3D_1^{\mathrm T}$ partial wave emphasizes the relevant role played in the 
calculations by tensor correlations, which are however present and important 
also in the other components. 

In Fig.~\ref{fig:fig2} the shapes of the recoil-momentum distributions for the 
different relative states are basically determined by the value of $L$, since
final-state interaction modifies the momentum of the pair but does not change 
drastically the shape. The shape of the final cross section is driven by the 
component which gives the major contribution, i.e. for the $1^+_1$ state and 
in the considered kinematics by $^3S_1$. An important, although less relevant 
role, is also played by $^3D_1^{\mathrm T}$, mainly at large values of 
$p_{\mathrm{B}}$, while $^3D_1$ is less important and $^1P_1$ is practically 
negligible. The final cross section is thus a combination of states with 
$L= 0$ and $L= 2$. For low values of $p_{\mathrm{B}}$ it has a typical 
$s$-wave shape, while at large values of $p_{\mathrm{B}}$ the $d$-wave 
contribution prevails.

\begin{figure}
\epsfysize=15.0cm
\begin{center}
\makebox[16.4cm][c]{\epsfbox{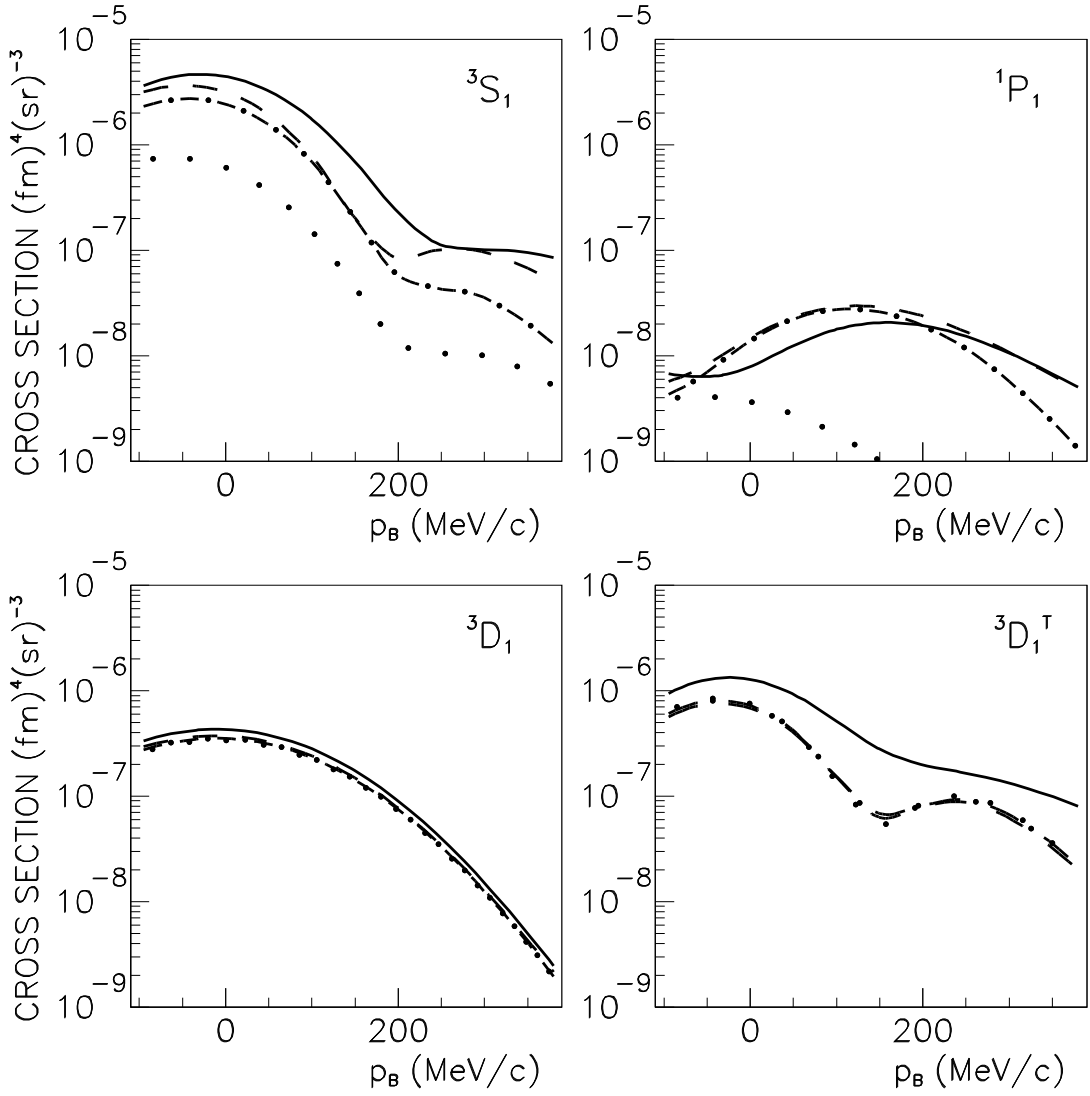}}
\end{center}
\caption[]{The differential cross section  of the $^{16}$O(e,e$'$pn)
reaction as a function of the recoil momentum $p_{\mathrm{B}}$ for the same 
transition and in the same kinematics as in Fig.~\ref{fig:fig1}. Optical
potential as in Fig.~\ref{fig:fig1}. Separate contributions of different
partial waves of relative motions are drawn: $^3S_1$, $^1P_1$, $^3D_1$ 
indicates the component already present in the uncorrelated wave function, 
$^3D_1^{\mathrm T}$  the component due to tensor correlations. The 
dotted lines give the separate contribution of the one-body current, the 
dot-dashed lines the sum of the one-body and seagull currents, the dashed 
lines the sum of the one-body, seagull and pion-in-flight currents and the 
solid lines the total result, where also the contribution of the 
$\Delta$-current is added.  
\label{fig:fig3}
}
\end{figure}

The dependence of the cross sections calculated for different relative states 
on the various components of the nuclear current is displayed in 
Fig.~\ref{fig:fig3}. Two-body currents and correlations play a different role 
in $^3S_1$ and $^3D_1$ knockout. The seagull and $\Delta$-isobar currents give 
the major contribution to $^3S_1$. Here the one-body current is not too 
relevant, although non negligible. In contrast, it is much more important in 
$^3D_1$, where it is dominant, and in $^3D_1^{\mathrm T}$, where the 
contributions of the one-body and $\Delta$ currents are of about the same size 
and add up in the cross section. Two-body currents give the major contribution 
also to $^1P_1$, but the role of this relative wave in the cross section is 
very small. 

The major contribution to the total cross section is thus given by $^3S_1$ and 
to a lesser extent also by $^3D_1^{\mathrm T}$. The relevant role played by 
correlations in $^3D_1^{\mathrm T}$, which is determined by tensor 
correlations, and in $^3D_1$ is therefore the main responsible for the 
important effect of the one-body current in the cross section of 
Fig.~\ref{fig:fig1}.

\begin{figure}
\epsfysize=9.0cm
\begin{center}
\makebox[16.4cm][c]{\epsfbox{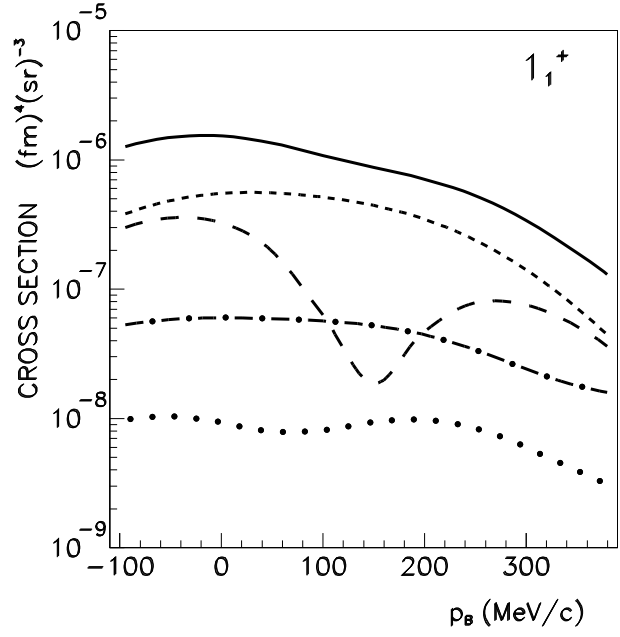}}
\end{center}
\caption[]{The same as Fig.~\ref{fig:fig1} but with the simpler approach of
Ref.~\cite{gnn} for the two-nucleon overlap. The single-particle wave functions
are taken from Ref.~\cite{ES} and the correlation function from Ref.~\cite{GD}.
Line convention as in  Fig.~\ref{fig:fig1}. 
\label{fig:fig4}
}
\end{figure}

In Fig.~\ref{fig:fig4} the same quantities as in Fig.~\ref{fig:fig1} are shown,
but the two-nucleon overlap has been calculated with the simpler prescription 
of ref.~\cite{gnn}, i.e. by the product of the pair function of the shell 
model, described for $1^+_1$ as a pure ($p_{1/2}$)$^{-2}$ hole, and of a 
Jastrow type central and state independent correlation function. 

The large differences between the cross sections in Figs.~\ref{fig:fig1} 
and~\ref{fig:fig4} indicate that a refined description of the two-nucleon 
overlap, involving a careful treatment of both aspects related to nuclear 
structure and NN correlations, is needed to give reliable predictions of the 
size and the shape of the (e,e$'$pn) cross section. This result, that is well 
established for the (e,e$'$pp) reaction also in comparison with 
data~\cite{NIKHEF,Rosner,klas1}, is here confirmed also for the (e,e$'$pn) 
reaction. In particular, the calculated cross sections are very sensitive to 
the treatment of NN correlations. The difference between the results in 
Figs.~\ref{fig:fig1} and~\ref{fig:fig4} is indeed dramatic for the separate 
contribution of the one-body current, which is completely determined by 
correlations. This contribution, that in Fig.~\ref{fig:fig1} is competitive 
with or even larger than that of the two-body current, is negligible in 
Fig.~\ref{fig:fig4}. The correlation function adopted in Fig.~\ref{fig:fig4} 
neglects the difference between $pp$ and $np$ correlations and their state 
dependence. Moreover, it does not take into account tensor correlations, which 
are crucial in the $pn$ channel. Therefore, this treatment represents a too 
simple prescription, that is not able to give 
a proper description of correlations effects in the (e,e$'$pn) cross section. 
The contribution of the two-body current, which is expected to be generally 
very important or even dominant in $pn$ knockout, is also expected to be much 
less affected by correlations and thus by their treatment. The comparison 
between Figs.~\ref{fig:fig1} and~\ref{fig:fig4} indicate, however, that the 
model used for the two-nucleon overlap function has meaningful effects also on 
the various terms of the two-body current. These effects are of course less 
dramatic than those obtained on the one-body current, but are anyhow large. 

The comparison between Figs.~\ref{fig:fig1} and~\ref{fig:fig4} is only an
example, but it clearly shows that reliable numerical predictions of the
(e,e$'$pn) cross section require a careful description of the two-nucleon
overlap functions involving a refined and consistent treatment of long-range and
short-range correlations. 

\begin{figure}
\epsfysize=9.0cm
\begin{center}
\makebox[16.4cm][c]{\epsfbox{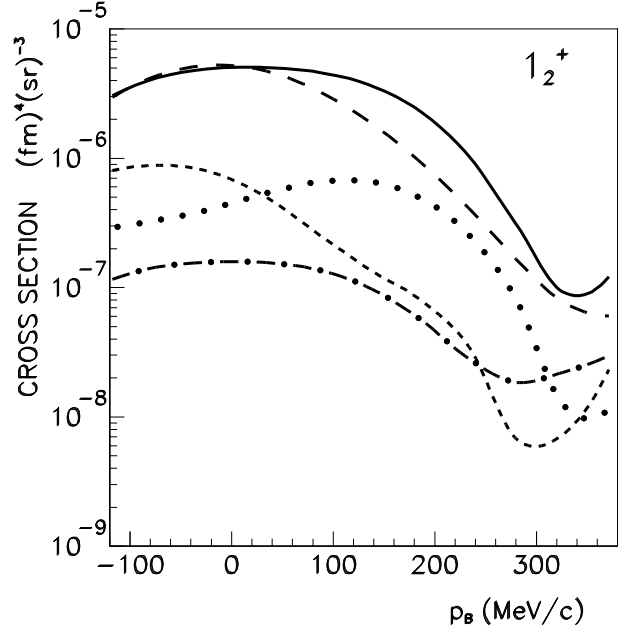}}
\end{center}
\caption[]{The same as Fig.~\ref{fig:fig1} for the transition to the $1^+_2$ 
state of $^{14}$N ($E_{2\mathrm{m}} = 26.91$ MeV). 
\label{fig:fig5}
}
\end{figure}

\begin{figure}
\epsfysize=14.0cm
\begin{center}
\makebox[16.4cm][c]{\epsfbox{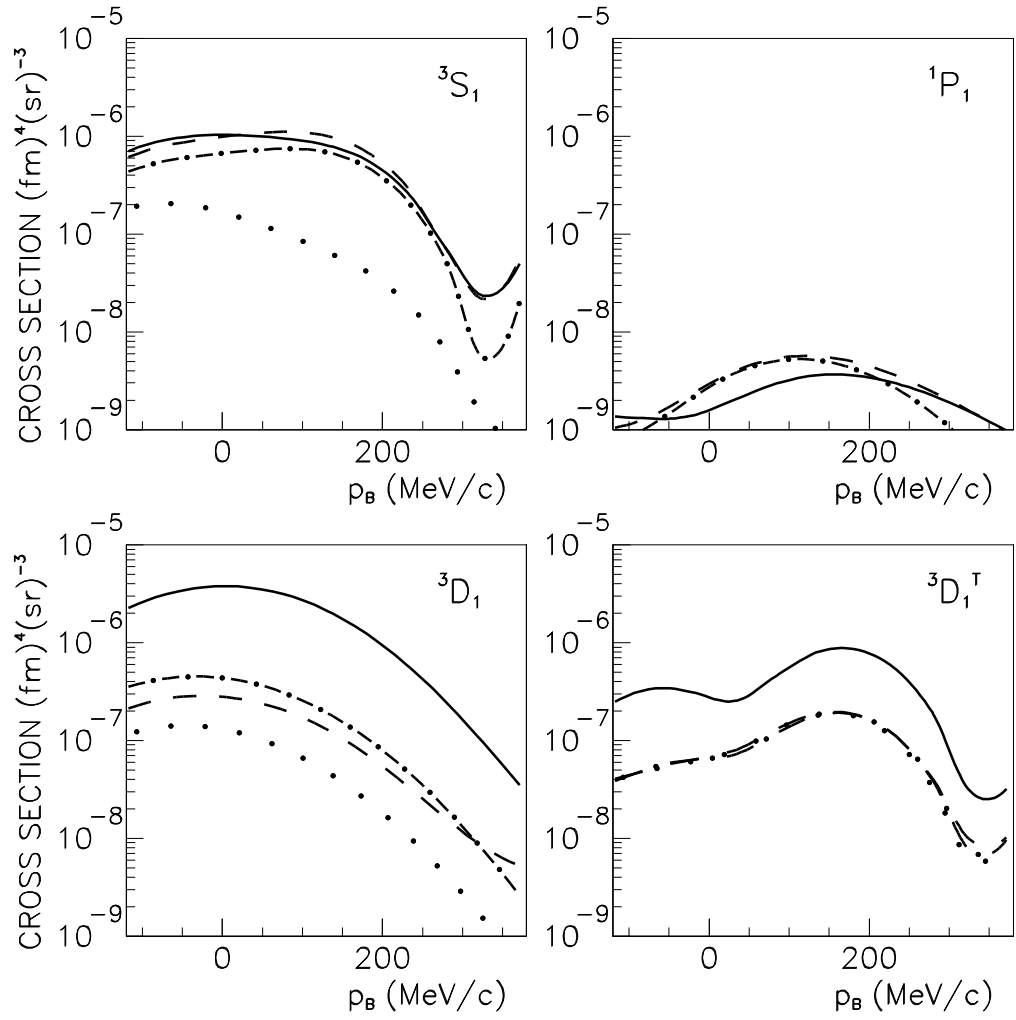}}
\end{center}
\caption[]{The same as Fig.~\ref{fig:fig3} for the transition to the $1^+_2$ 
state of $^{14}$N. 
\label{fig:fig6}
}
\end{figure}

The cross sections for the transition to the 1$^+_2$ state are displayed in 
Figs.~\ref{fig:fig5} and~\ref{fig:fig6}. The two-nucleon overlap function for 
this state contains the same components in terms of relative and c.m. wave 
functions and the same defect functions as for the 1$^+_1$ ground state, but
they are weighed with different amplitudes $a^{J}_{\nu_1\nu_2}$ in 
Eq.~(\ref{eq:ppover}). In practice the two overlap functions have different 
amplitudes for $p_{1/2}$ and $p_{3/2}$ holes (see Table I). This has the 
consequence that the cross 
sections in Figs.~\ref{fig:fig1} and~\ref{fig:fig5} have a different shape and 
are differently affected by the various terms of the nuclear current. The shape 
of the recoil-momentum distribution in Fig.~\ref{fig:fig5} indicates a more 
prominent role of the components with c.m. $L= 2$. The comparison between the 
contributions of the different terms of the nuclear current in the two states 
shows that the $\Delta$-current, which in Fig.~\ref{fig:fig1} for 1$^+_1$ is 
much less important than the one-body and seagull terms, is much larger and 
even dominant in the 1$^+_2$ state. In contrast, the one-body and seagull 
currents, that play the main role for the transition to the ground state, are
less important for 1$^+_2$. The one-body current gives a meaningful 
contribution only for recoil momenta between 100 and 200 MeV/$c$. The
contribution of the seagull current is generally small and that of the 
pion-in-flight current is even smaller. This result can 
better be understood in Fig.~\ref{fig:fig6}, where separate contributions of 
the relative waves and their dependence on the terms of the nuclear current are
displayed. There are two main differences between the cross sections in 
Figs.~\ref{fig:fig3} and~\ref{fig:fig6}: the different role played by $^3S_1$ 
and $^3D_1$, and the dominant contribution of the $\Delta$-current in $^3D_1$
knockout. The $^3D_1$ relative wave function in Fig.~\ref{fig:fig6} is much 
more important than for the transition to the 1$^+_1$ state and gives the major
contribution to the cross section. However, while in the corresponding 
situation for the 1$^+_1$ state it is dominated by the one-body current, in 
1$^+_2$ it is dominated by the $\Delta$-current. Moreover, the $\Delta$-current
is for this transition more important also in $^3D_1^{\mathrm T}$ knockout. It 
is less meaningful in $^3S_1$, but the role of this relative state in the final
cross section is less relevant for the 1$^+_2$ than for the 1$^+_1$ state. 
Thus, the major contribution to the final cross section is given by the 
$\Delta$-current. 

All the three states $^3S_1$, $^3D_1$ and $^3D_1^{\mathrm T}$ play a 
significant role in the cross section, while $^1P_1$ is negligible 
also for the 1$^+_2$ state. The comparison between the shapes of the recoil
momentum distributions of $^3S_1$ knockout in Figs.~\ref{fig:fig3} 
and~\ref{fig:fig6} indicates a more prominent role of the components with c.m. 
$L= 2$ for the 1$^+_2$ state. This feature, which is more evident at large 
values of the recoil momentum, explains the shape of the final cross section in 
Fig.~\ref{fig:fig5}. 

\begin{figure}
\epsfysize=9.0cm
\begin{center}
\makebox[16.4cm][c]{\epsfbox{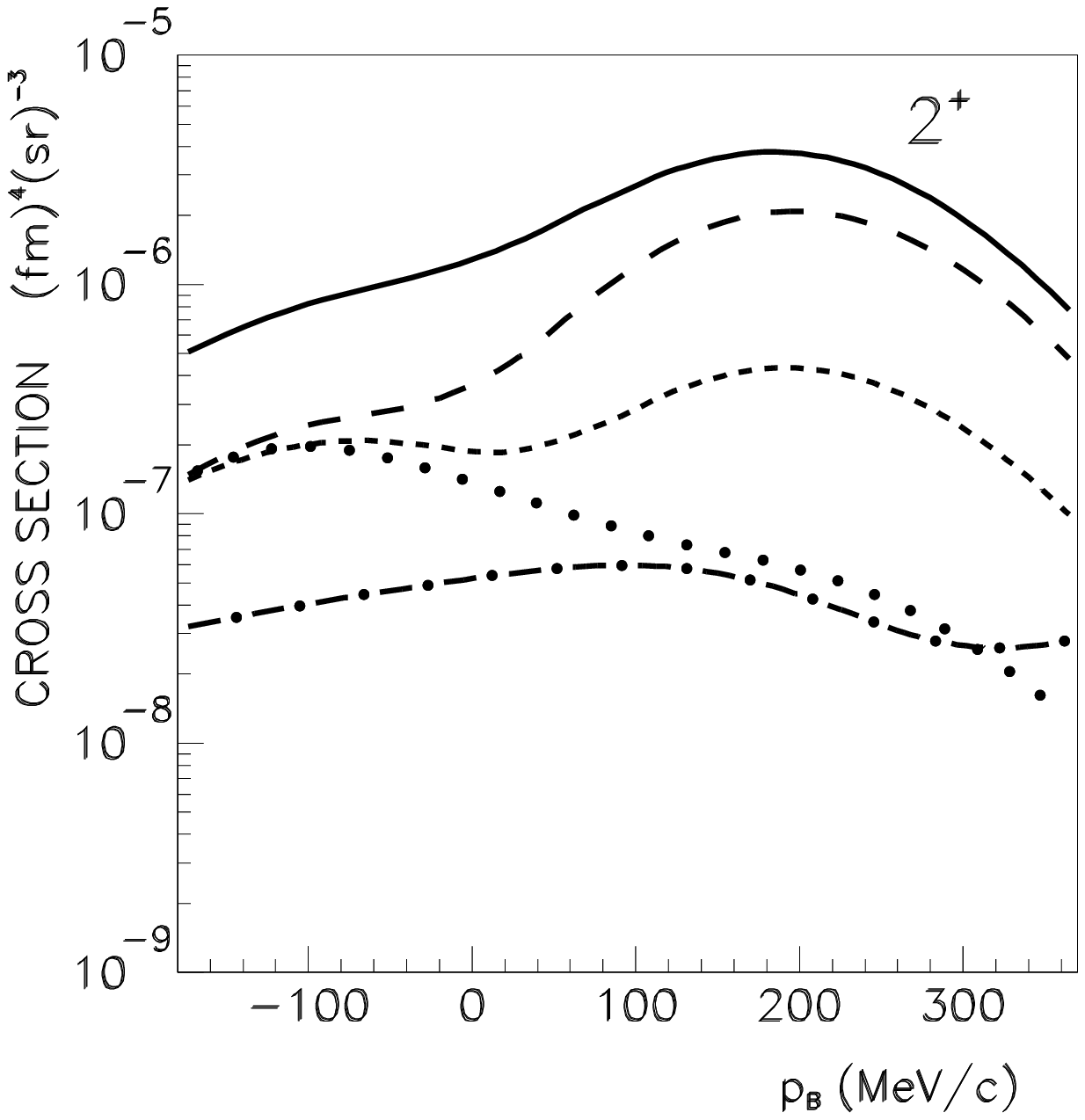}}
\end{center}
\caption[]{The same as Fig.~\ref{fig:fig1} for the transition to the $2^+$ 
state of $^{14}$N ($E_{2\mathrm{m}} = 29.99$ MeV). 
\label{fig:fig7}
}
\end{figure}

\begin{figure}
\epsfysize=15.0cm
\begin{center}
\makebox[16.4cm][c]{\epsfbox{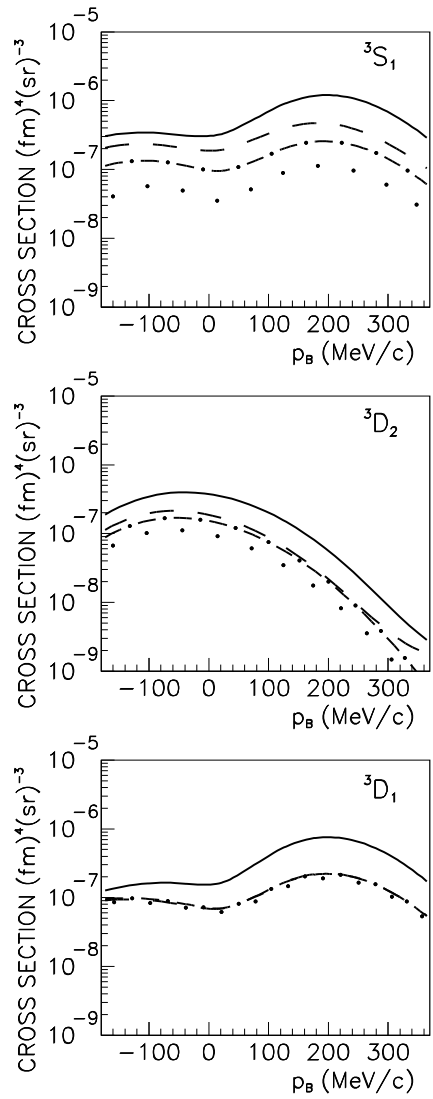}}
\end{center}
\caption[]{The differential cross section  of the $^{16}$O(e,e$'$pn)
reaction as a function of the recoil momentum $p_{\mathrm{B}}$ for the same 
transition and in the same kinematics as in Fig.~\ref{fig:fig7}. The optical
potential is taken from Ref.~\cite{Nad}. Separate contributions of different
partial waves of relative motions are drawn: $^3S_1$, $^3D_2$, $^3D_1$.
Line convention as in Fig.~\ref{fig:fig3}  
\label{fig:fig8}
}
\end{figure}

The results for the transition to the 2$^+$ state are displayed in 
Figs.~\ref{fig:fig7} and~\ref{fig:fig8}. The shape of the recoil-momentum 
distribution for this transition is different from that of the other states 
previously considered. The position of the maximum  at
$p_{\mathrm{B}}\simeq 200$ MeV/$c$ is an indication of the relevant 
contribution of the partial wave functions with c.m. $L= 2$. The 
$\Delta$-current dominates with respect to the other terms in the maximum 
region, whereas at low values of the recoil momentum all the three terms, 
$\Delta$ seagull and one-body, contribute with similar weights. The 
contribution of the pion-in-flight current is always very small. 

The separate cross sections produced by different partial waves of relative 
motion are displayed in Fig.~\ref{fig:fig8}. For this final state they are: 
$^3S_1$, which is combined with $L= 0$ and $L= 2$, $^3D_2$, combined with 
$L= 0$, and $^3D_1$, not present in the uncorrelated wave function and 
produced by tensor correlations, which is also combined with $L= 0$ and $L= 2$.
The components with $L= 0$ prevail at low values of $p_{\mathrm{B}}$, where
they give a contribution of about the same size for all the three relative 
states. The components with $L= 2$, present only in $^3D_1$ and $^3S_1$, are 
dominant at large values of the recoil momentum. In $^3S_1$ all the different 
terms of the nuclear current, including pion-in-flight, give a meaningful 
contribution to the cross section. In $^3D_2$ the role of pion-in-flight is 
negligible, but the other terms all give a meaningful contribution. In $^3D_1$ 
the cross sections calculated with the one-body and $\Delta$ currents are of 
about the same size, whereas those calculated with seagull and pion-in-flight 
are much smaller. The $\Delta$-current is therefore important in all the three 
relative states, in particular at large values of the recoil momentum, while 
the one-body current gives a more meaningful contribution to $^3D_1$ and 
$^3D_2$ knockout and at low values of $p_{\mathrm{B}}$.

\section{Summary and conclusions}

The cross section has been calculated for the $^{16}$O(e,e$'$pn)$^{14}$N 
reaction leading to the ground state and the first excited $1^+$ and $2^+$
states in the residual nucleus $^{14}$N. The calculations account for
correlation effects, using correlated wave functions calculated in the framework
of the coupled cluster method, MEC contributions and the effects of final state
interaction of the outgoing nucleons with the residual nucleus. As an example we
consider the so-called super-parallel kinematics with a value for the energy
($\omega$ = 215 MeV) and momentum transfer ($q$ = 316 MeV/c) which is 
appropriate for experiments to be performed at MAMI in Mainz.
The cross sections for (e,e$'$pn) are about an order of magnitude larger than 
corresponding cross sections for (e,e$'$pp) reactions. This enhancement is 
partly due to the importance of MEC, the pion seagull contribution in 
particular, and partly due to the enhanced importance of correlation effects in 
$pn$ pairs as compared to $pp$ pairs. The relative importance of the MEC 
contribution as compared to the correlation effects depends on the final state 
of the residual nucleus. This is similar to the observations made for 
(e,e$'$pp)~\cite{klas1}. The calculated cross sections are
rather sensitive to details of the nuclear correlations considered
and in particular to the presence of the tensor component. In fact, 
a correlation function without this component gives quite 
different results for the calculated cross sections. Also the choice of the 
structure amplitudes is important and produces a sensible effect in
the shape and size of the cross sections. This means that an accurate 
determination of the spectral density and therefore of the overlap 
wave functions is required in order to predict in a satisfactory 
and reliable way the experimental yields. This will
hopefully  enable us to disentangle the effects of the MEC  
contributions from those of the nuclear correlations.
These observations support the efforts to complete our knowledge on nuclear
correlations obtained from (e,e$'$pp) reactions by corresponding experiments 
on (e,e$'$pn).

These investigations have partly been supported by the Deutsche
Forschungsgemeinschaft (``Schwerpunktprogramm, WA 728/3'').


\begin{table}
\begin{tabular}{|cccccc|}
& $J^\pi$ &\, ($p_{3/2}$)$^{-2}$   & \,\,\,\,\,\, ($p_{3/2}p_{1/2}$)$^{-1}$   
&\, ($p_{1/2}$)$^{-2}$ & \\
\hline
& 1$^+_1$   & $0.070$ & $-0.455$   &  $ 0.606$ &\\
& 1$^+_2$   & $0.271$ & $-0.544$   &  $-0.460$ &\\
& 2$^+  $   & $0.   $ & $ 0.765$   &  $ 0.   $ &\\
\end{tabular}

\bigskip
 
\caption[Table I]{
The coefficients $a^{J}_{\nu_1\nu_2}$ in Eq.~(\ref{eq:ppover}) for the 
low-lying states states of $^{14}$N.
\label{tab:pn}
}
\end{table}
\vfil
\vfil\eject

\vfil 
\vfil\eject
\vfil
\vfil\eject
\vfil
\vfil\eject

\vfil\eject
\vfil
\vfil\eject
\vfil
\vfil\eject
\end{document}